\documentclass[aps,pre,twocolumn,showpacs,superscriptaddress]{revtex4}
\usepackage{amssymb}
\usepackage{graphicx}

\begin{document}

\title{Reachability and recoverability of sink nodes in growing acyclic directed
networks}

\author{Valmir C. Barbosa}
\affiliation{Programa de Engenharia de Sistemas e Computa\c c\~ao, COPPE,
Universidade Federal do Rio de Janeiro,
Caixa Postal 68511, 21941-972 Rio de Janeiro - RJ, Brazil}

\begin{abstract}
We study the growth of networks from a set of isolated ground nodes by the
addition of one new node per time step and also of a fixed number of directed
edges leading from the new node to randomly selected nodes already in the
network. A fixed-width time window is used so that, in general, only nodes that
entered the network within the latest window may receive new incoming edges. The
resulting directed network is acyclic at all times and allows some of the ground
nodes, then called sinks, to be reached from some of the non-ground nodes. We
regard such networks as representative of abstract systems of partially ordered
constituents, for example in some of the domains related to technological
evolution. Two properties of interest are the number of sinks that can be
reached from a randomly chosen non-ground node (its reach) and, for a fixed
sink, the number of nonoverlapping directed paths through which the sink can be
reached, at a given time, from some of the latest nodes to have entered the
network. We demonstrate, by means of simulations and also of analytic
characterizations, that reaches are distributed according to a power law and
that the desired directed paths are expected to occur in very small numbers,
perhaps indicating that recovering sinks late in the process of network growth
is strongly sensitive to accidental path disruptions.
\end{abstract}

\pacs{89.75.Hc, 05.65.+b, 89.75.Da, 89.75.Fb}

\maketitle

\section{Introduction}

The study of large, essentially unstructured networks of interacting elements,
also referred to as complex networks, has in the past several years received
considerable attention. The main motivation behind so much interest has been
the realization that networks occurring in many natural, technological, and
social domains have common statistical properties that, though governed by
strictly local interactions among the networks' elements, relate globally to the
networks' structure or functionality. A comprehensive collection of papers
spanning the main aspects of this emerging discipline, from origins to
representative applications, can be found in \cite{bs03,nbw06}.

While it seems correct to say that most network models studied so far are
undirected, reflecting the fact that the local interactions occur between pairs
of interconnected elements in any of the two possible directions (this is the
case, for example, of the networks that represent the Internet at some level),
there are also several cases in which interactions are inherently
unidirectional, as for example the WWW \cite{baj00}, networks of bibliographic
citations \cite{v01}, and also networks that arise from certain flows of
information in computer networks \cite{sb06a,sb06b,sb07}. Unidirectional
interactions give rise to directed networks (that is, networks whose edges have
directions), which in turn have been studied for both structural
\cite{dms01,bds03,bds04,bm06} and functional \cite{kkk02,g04} properties.

The structure of directed networks is considerably more intricate than that of
undirected networks, and this is due primarily to the existence of directed
cycles, that is, node sequences in which it is possible to return to any node
by following edges along their directions. The existence of such cycles in a
directed network is strictly necessary for nontrivial strongly connected
components to appear, so it comes as no surprise that many of the network's
properties depend on whether directed cycles exist, how large they are, and how
they relate to other structures in the network. So, even though some attention
has been given to network elements that lie outside directed cycles \cite{m03}
or to how the network looks when directed cycles are broken \cite{c04}, a fair
appraisal seems to be that studying directed networks has so far concentrated
primarily on properties that depend on the existence of directed cycles.

However, we find that a surprising number of systems are naturally representable
by directed networks that are intrinsically acyclic, that is, contain no
directed cycles (even though plenty of cycles exist if one ignores the edges'
directions). Such networks exist at much more abstract levels than the majority
of the networks that have received attention from researchers, reflecting in
general the partial order that is inherent to their nature or to the manner in
which they are constructed. Important examples are: networks of immediate event
precedence, both in history \cite{s96} and in the unfolding of distributed
computations \cite{l78}; networks of object inheritance in object-oriented
programs \cite{tcsh06}; the probabilistic graphical models, known as Bayesian
networks, that represent the causal relationships among random variables in some
artificial-intelligence systems \cite{p88}; networks that represent possible
deductions in axiomatic systems of formal proof \cite{c02}; and networks of word
etymology in large language groups \cite{babel-url}.

Perhaps the reason why systems such as these have not yet been approached from
a complex-network perspective is ultimately the elusiveness that they have about
them. In some cases, data are simply not readily obtainable, as seems to be the
case of the networks that reflect the innards of large software or
artificial-intelligence systems. In others, as in the history and etymology
systems, even defining the network's elements depends on data that are no longer
extant and thus requires extensive hypothesizing. Even so, it seems possible to
postulate some prototypical growth model for acyclic directed networks and then
use it in the study of properties that are expected to be of interest.

Our approach in this paper is to study the growth of acyclic directed networks
from an initial set of ground nodes by the continual addition of new nodes and
directed edges. At each time step, the growth is limited to the addition of one
single node and a fixed number of edges outgoing from that node to randomly
selected nodes already in the network. We impose a constraint on which are the
nodes toward which new edges may be added: as a new node enters the network, the
outgoing edges it acquires must necessarily lead to nodes inside a fixed-size
window representing that time step's immediate past. Both finite and infinite
windows are considered, so we hope to be contemplating a wide variety of
circumstances in regard to the previously mentioned networks as well as others.

Unlike most other studies of complex networks, in the present case the central
entities to be observed are not node degrees (distributions are trivially
obtainable for both in- and out-degrees, as we discuss shortly), but have to do
instead with whether (and from which nodes) the ground nodes remain reachable as
time elapses and, if they do, the nature of the directed paths that lead to
them. What we have found is that ground-node reachability depends on how the
number of ground nodes relates to window size, and also that the number of
ground nodes that can be reached is at times distributed as a power law. As for
recovering ground nodes from the latest nodes added to the network, this is
expected to be achievable only through a very small number of nonoverlapping
directed paths, thus indicating high susceptibility to failure should one such
path be disrupted.

\section{The model and basic properties}

We study network evolution for discrete time $t\ge 1$ from an initial set of
$n_0$ isolated ground nodes. One new node is added per time step, so the
elapsing of time step $t$ causes the network to have $n_0+t$ nodes. We identify
the ground nodes by the nonpositive integers $-n_0+1,\ldots,0$, thus imposing an
arbitrary order on them, even though they are all assumed to be present when
network growth begins. We also use $t$, interchangeably, to refer both to time
step $t$ and to the node added at that time step. Upon entering the network,
node $t$ acquires two outgoing edges leading to distinct nodes chosen randomly
from the set $\{\max\{-n_0+1,t-w\},\ldots,t-1\}$ for some window $w\ge n_0$. If
$t\le w+1$, then this set contains
\begin{equation}
w_t=\min\{n_0,w-t+1\}
\end{equation}
ground nodes; it contains no ground nodes otherwise. [Note that the choice of
$2$, as opposed to some other constant, as the number of outgoing edges per node
added to the network is qualitatively irrelevant, so we make it for simplicity's
sake only. Similarly, we rule out the possibility of $w<n_0$, because this is
qualitatively equivalent to using a number of ground nodes equal to $w$ (since
it implies that $n_0-w$ ground nodes are guaranteed to remain isolated
indefinitely).]

Every non-ground node has an out-degree of exactly $2$. As for in-degrees, we
may concentrate on some non-ground node $i$ and let $k\in\{0,\ldots,w\}$. The
probability that $i$ has in-degree $k$ is clearly given by
\begin{equation}
{{w}\choose{k}}
\left(\frac{2}{w}\right)^k
\left(1-\frac{2}{w}\right)^{w-k}
\approx\frac{2^ke^{-2(1-k/w)}}{k!},
\end{equation}
which approximates the probability that, at time $t\gg n_0$, a randomly chosen
node has in-degree $k$. For $k\ll w$, it approaches the mean-$2$ Poisson
distribution. (Note that, if we condition on ground nodes exclusively, the
in-degree distribution becomes more concentrated at low degrees than the
mean-$2$ Poisson, which implies a lower mean value.)

We henceforth refer to every non-isolated node having no outgoing edges as a
sink, and to every non-isolated node having no incoming edge as a source.
Clearly, every ground node becomes a sink when picked to be directed an edge at
for the first time, and conversely only ground nodes may be sinks. Likewise,
every non-ground node is a source upon entering the network, though it may cease
being one afterward; conversely, no ground node may be a source.

Let $S_t$ denote the expected number of sinks just before the addition of node
$t$ to the network. We have $S_1=0$ and, for $t\ge 1$,
\begin{equation}
S_{t+1}=S_t+\Delta_t,
\label{eq:difference}
\end{equation}
where $\Delta_t$ is the expected number of new sinks created when node $t$ is
added. Of the $w_t$ ground nodes that may acquire a new incoming edge at time
$t$, let those that are already sinks amount to an expected number $f_t$. Then
$f_t=(w_t/n_0)S_t$ and $w_t-f_t=w_t(1-S_t/n_0)$.

The number of node pairs from which to choose at time $t$ is
$(w_t+t-1)(w_t+t-2)/2$. Of these, $[w_t+t-1-(w_t-f_t)](w_t-f_t)$ are expected to
lead to the creation of one new sink, while $(w_t-f_t)(w_t-f_t-1)/2$ others are
expected to lead to the creation of two new sinks. We then obtain
\begin{eqnarray}
\Delta_t
&=&\frac{2(w_t-f_t)(f_t+t-1)}{(w_t+t-1)(w_t+t-2)}\nonumber\\
&&\hspace{0.5in}\mbox{}+\frac{2(w_t-f_t)(w_t-f_t-1)}{(w_t+t-1)(w_t+t-2)}\\
&=&\frac{2w_t(1-S_t/n_0)}{w_t+t-1}.
\end{eqnarray}
Approximating (\ref{eq:difference}) by a differential equation yields two
possibilities, depending on $t$. For $1\le t\le w+1-n_0$, $w_t=n_0$ and we get
\begin{equation}
\frac{dS_t}{dt}+\frac{2S_t}{n_0+t-1}=\frac{2n_0}{n_0+t-1},
\end{equation}
thence
\begin{equation}
S_t=\frac{n_0(t-1)(2n_0+t-1)}{(n_0+t-1)^2}
\label{eq:solution1}
\end{equation}
is obtained from $S_1=0$. For $w+1-n_0\le t\le w+1$, $w_t=w-t+1$ and we get
\begin{equation}
\frac{dS_t}{dt}+\frac{2(w-t+1)S_t}{wn_0}=\frac{2(w-t+1)}{w},
\end{equation}
thence
\begin{equation}
S_t=n_0\left\{1-\left(\frac{n_0}{w}\right)^2
\exp\left[\left(\sqrt\frac{w}{n_0}-\frac{t-1}{\sqrt{wn_0}}\right)^2
-\frac{n_0}{w}\right]\right\}
\label{eq:solution2}
\end{equation}
results from $S_{w+1-n_0}=n_0[1-(n_0/w)^2]$ [cf.\ (\ref{eq:solution1})]. Notice
that expressing $S_t/n_0$ as a function of $(t-1)/n_0$ in (\ref{eq:solution1}),
which is already independent of $w$, yields a constant with respect to $n_0$ as
well. Doing the same in (\ref{eq:solution2}) reveals an exclusive dependence on
the ratio $n_0/w$.

Beginning at $t=w+1$, it is no longer possible for any sink to be created, so
the expected number of sinks settles at the value, henceforth denoted by
$S(n_0/w)$, given by
\begin{equation}
S(n_0/w)=S_{w+1}=n_0\left[1-\left(\frac{n_0}{w}\right)^2e^{-n_0/w}\right],
\end{equation}
following (\ref{eq:solution2}). For $w=n_0$, this becomes $S(1)=n_0(1-e^{-1})$,
which limits the expected number of sinks at about $63.21\%$ of the ground
nodes. As $w$ grows, $S(n_0/w)$ approaches $n_0$ asymptotically.

Our study on the recoverability of sinks will be based on the nodes that, at
time $t$, remain sources inside the latest window (i.e., the window comprising
nodes $t-w+1,\ldots,t$). The probability that a node $i$ inside this window
remains a source through time $t$ is $[(w-2)/w]^{t-i}$. The expected number of
sources inside the latest window, denoted by $R$, is then
\begin{equation}
R=\sum_{i=t-w+1}^t\left(\frac{w-2}{w}\right)^{t-i}
\approx w\left(\frac{1-e^{-2}}{2}\right),
\end{equation}
amounting therefore to roughly $43.23\%$ of the nodes inside the window.

\section{Reachability and recoverability of sinks}

\subsection{Reachability}

At time $t$, we say that a ground node is reachable from one of the $n_0+t$
nodes of the network when a directed path exists between them leading to the
ground node. All ground nodes are reachable from themselves, but only sinks are
reachable from non-ground nodes. The reach of a node is the number of ground
nodes that are reachable from it. A node has unit reach if and only if it is a
ground node, and the reach of a non-ground node refers to sinks exclusively.

Let $P_t(r)$ be the probability that, at time $t$, a randomly chosen node has
reach $r$. Clearly,
\begin{equation}
P_t(1)=\frac{n_0}{n_0+t}.
\end{equation}
For $r>1$, however, we expect the number of sinks in the network to play a role
in defining the value of $P_t(r)$.

As a node enters the network and connects out to two previously existing nodes,
its reach has to account for every sink that is reachable from either of those
two nodes. In the relatively early stages of network formation, and for
sufficiently large $n_0$, it is likely that no sink is reachable from the two
nodes concomitantly, and in this case the new node's reach is simply the sum of
their reaches. This becomes progressively less likely later on in the evolution
of the network, thus making accurate predictions of $P_t(r)$ very difficult.

\begin{figure}
\centering
\includegraphics[scale=0.45]{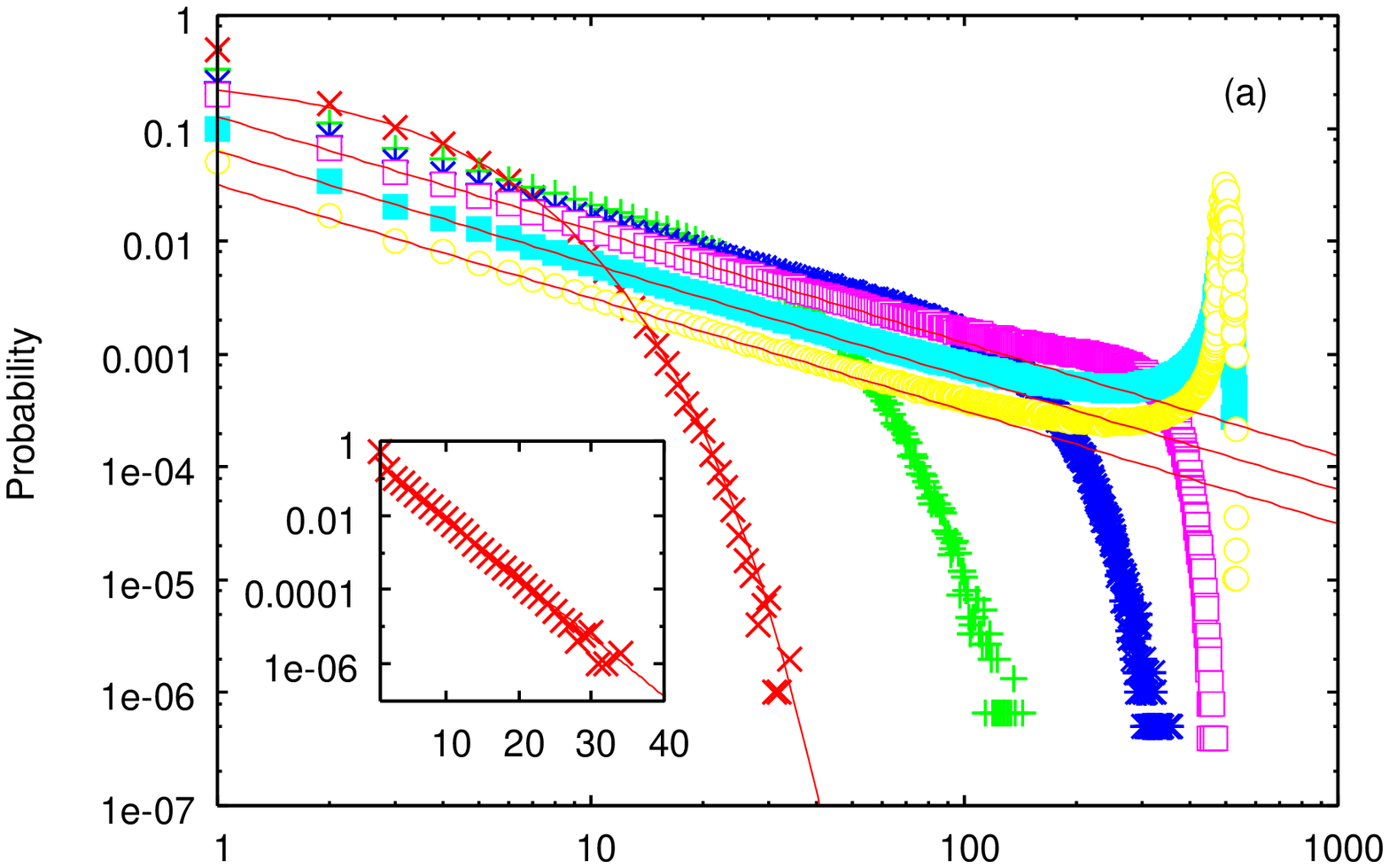}\\
\includegraphics[scale=0.45]{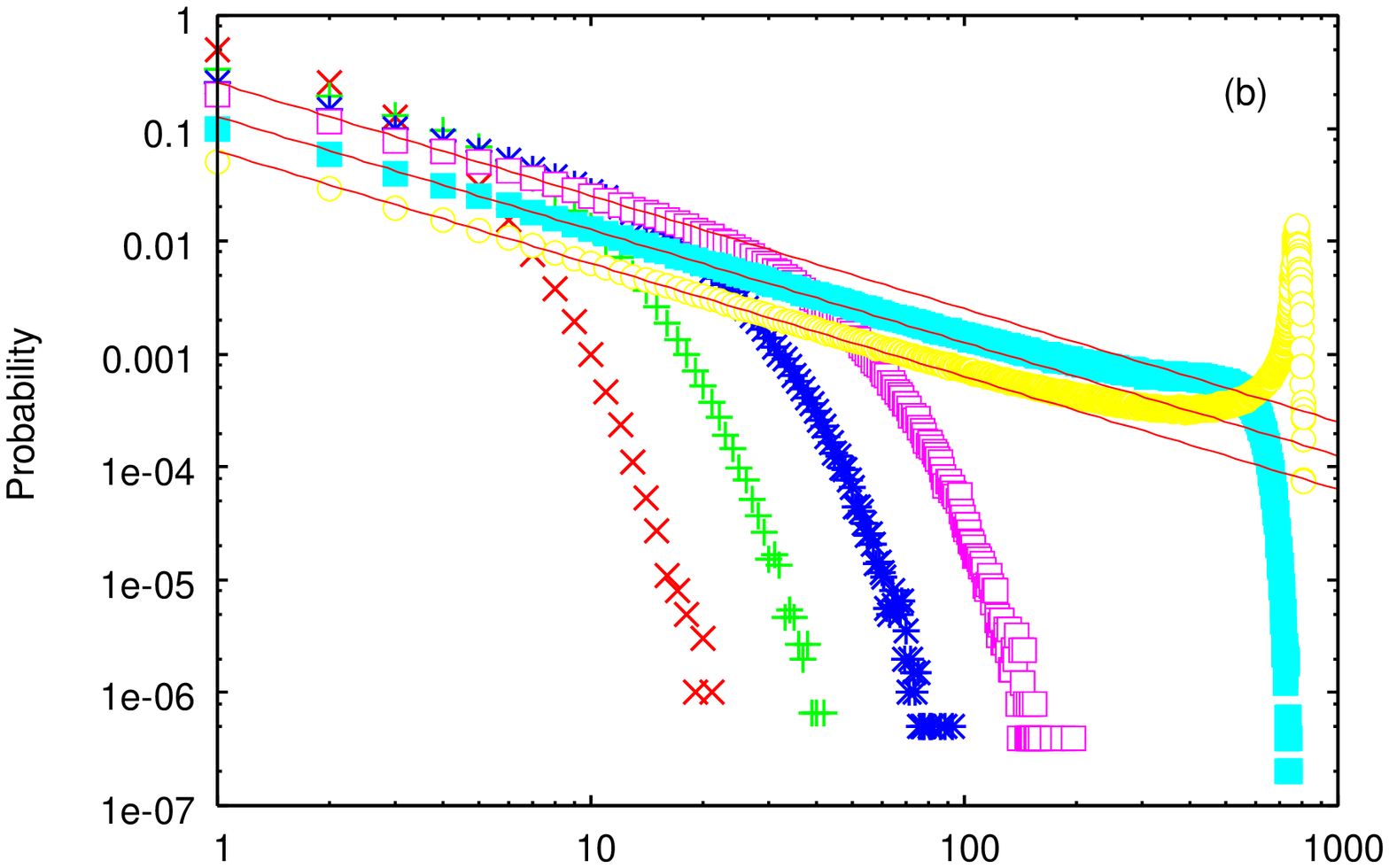}\\
\includegraphics[scale=0.45]{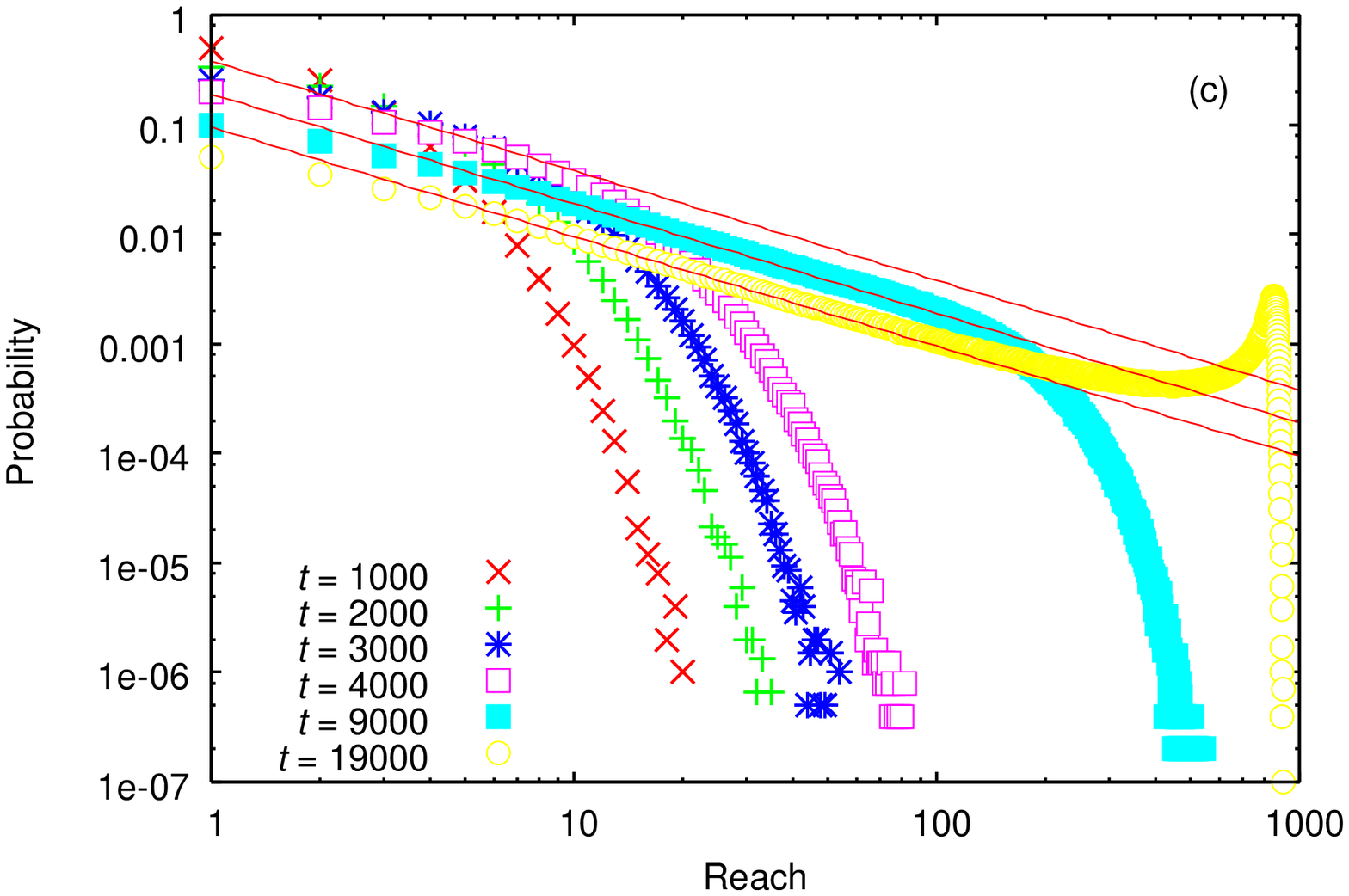}
\caption{(Color online) Reach distribution for $n_0=1\,000$, with $w=n_0$ (a),
$w=2n_0$ (b), and $w=3n_0$ (c). Solid lines give the analytic predictions of
(\ref{eq:exp1}) and (\ref{eq:pl1}) for part (a), of (\ref{eq:pl2}) for parts (b)
and (c). All simulation data are averages over $500$ independent runs.}
\label{fig:wpl}
\end{figure}

Our finds regarding $P_t(r)$ are summarized in Figure~\ref{fig:wpl}, whose part
(a) refers to $w=n_0$. In this case we see that, initially, non-unit reaches
tend to be distributed exponentially. For $t=w=n_0$, in particular, the
exponential character of the distribution is very clear [cf.\ the inset in part
(a) of the figure] and may be expressed as
\begin{equation}
P_{n_0}(r)
\approx\left(\frac{S(1)}{2n_0}\right)a^r
=\left(\frac{1-e^{-1}}{2}\right)a^r,
\label{eq:exp1}
\end{equation}
for some constant $a$ such that $0<a<1$. Since the exponential seems to hold
across all pertinent reach values, we can find $a$ by requiring
\begin{equation}
P_{n_0}(1)+\sum_{r\ge 2}P_{n_0}(r)
=\frac{1}{2}+\left(\frac{1-e^{-1}}{2}\right)\sum_{r\ge 2}a^r
=1,
\end{equation}
which leads to $a\approx 0.6958$. It also seems that an exponential
approximation continues to hold for somewhat larger values of $t$. For $t\gg w$,
though, we expect more and more nodes of reach around $S(1)$ to appear, owing to
the finiteness of $w$. This is indeed what happens, but aside from this effect
we have also found that the passage of time leads the initial exponential
approximation to $P_t(r)$ to gradually become
\begin{equation}
P_t(r)
\approx\left(\frac{S(1)}{n_0+t}\right)r^{-1}
=\left(\frac{n_0(1-e^{-1})}{n_0+t}\right)r^{-1},
\label{eq:pl1}
\end{equation}
similar therefore to the power law known as Zipf's law.

As we increase $w$ beyond $n_0$ to $w=2n_0$ and $w=3n_0$, we obtain a similar
evolution of $P_t(r)$ with respect to $t$, including the progressive probability
accumulation around $r=S(1/2)$ or $r=S(1/3)$, depending on the case. This is
illustrated, respectively, in parts (b) and (c) of Figure~\ref{fig:wpl}, where
we see that the power-law regime is established only for increasingly larger
values of $t$. When this happens, a good approximation to $P_t(r)$ seems to be
\begin{equation}
P_t(r)
\approx\left(\frac{S(1)}{n_0+t}\right)\frac{r^{-1}}{n_0/w}
=\left(\frac{w(1-e^{-1})}{n_0+t}\right)r^{-1},
\label{eq:pl2}
\end{equation}
where, curiously, it is still $S(1)$ [not $S(1/2)$ or $S(1/3)$, as we might
expect] that drives the distribution, after the simple scaling by $n_0/w$.

\subsection{Recoverability}

We now examine the network's structure as it relates to the existence of
directed paths from the sources in $\{t-w+1,\ldots,t\}$, at time $t$, to the
sinks. While the average number of distinct paths over all such source-sink
pairs is distributed quite widely, when we look at paths that are not merely
distinct but edge-disjoint the situation is very different. For a given source
and a given sink, a group of directed paths between them is edge-disjoint if no
two paths in the group have any edges in common. The appropriate framework in
which to compute the maximum number of edge-disjoint directed paths between two
nodes is that of network flows.

Given a directed network with nonnegative numbers associated with the edges (the
edges' capacities), and assuming that it has at least one source and one sink,
the maximum flow from a source to a sink is an assignment of numbers to the
edges (their flows) such that: no edge flow exceeds the edge's capacity; the
total flow coming into any node equals that leaving the node (except for the
source and the sink); and moreover no other assignment results in a greater net
flow coming into the sink. By a well-known result from the theory of network
flows (the max-flow min-cut theorem), the number of edge-disjoint directed paths
from the source to the sink is precisely the maximum flow from the source to the
sink under unit capacities \cite{amo93}.

In our present context, the number of edge-disjoint directed paths from any
given source to any given sink is at most the minimum between the source's
out-degree (equal to $2$) and the sink's in-degree (distributed, as we have
noted, such that the mean is less than $2$). So we know, beforehand, that the
expected average number of such paths, taken over all source-sink pairs of
interest, lies somewhere in the interval $[0,2]$. Computing this number is
expected to require $RS(n_0/w)$ maximum-flow computations for each network. We
have used the publicly available, efficient HIPR code of \cite{hipr-url} for
$n_0=1\,000$ and three different values of $w$.

For $w=n_0$, we have found from $10$ independent runs that the expected average
is $0.5024$ at $t=4\,000$, growing to the roughly stable value of $1.2402$ at
$t=9\,000$. For $w=2n_0$ and $w=3n_0$, stabilization occurs later. For
$t=4\,000$ and $t=19\,000$, the expected averages are, respectively, as follows:
$0.0316$ and $1.4598$ for $w=2n_0$, $0.0122$ and $1.5069$ for $w=3n_0$. A small
increase is then observed at stability as $w$ becomes larger.

Another pertinent indicator of the recoverability of sinks from sources in the
latest window at time $t$ is the number of edge-disjoint directed paths from any
of the sources to a given sink. Clearly, the expected average number of such
paths, taken over all sinks, is some number in the interval $[0,2R]$, since the
expected number of sources is $R$ and each has the potential of contributing two
paths. However, the sink's in-degree remains distributed with a less-than-$2$
mean, so it is very unlikely for an expected average significantly larger than
$2$ to turn up. As for calculating the desired number of paths in a given
network for a given sink, we note that, unlike the preceding case, a little
artifice is needed before a maximum-flow computation can be performed (since it
is unclear what the source is in such a computation). What we do is to add
another source to the network and make capacity-$2$ directed edges outgo from it
to all original sources. The combined number of edge-disjoint directed paths
from the original sources to the sink is the maximum flow from the new source to
the sink. For each network, we expect $S(n_0/w)$ maximum-flow computations to be
needed.

\begin{figure}
\centering
\includegraphics[scale=0.45]{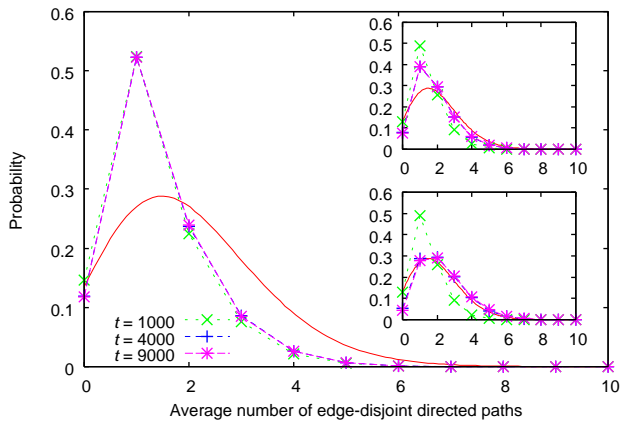}
\caption{(Color online) Distribution of the average number of edge-disjoint
directed paths from all sources to one sink for $n_0=1\,000$, with $w=n_0$,
$w=2n_0$ (top inset), and $w=3n_0$ (bottom inset). Solid lines give the mean-$2$
Poisson distribution. All simulation data are averages over $500$ independent
runs.}
\label{fig:npp}
\end{figure}

Results for this second indicator are shown in Figure~\ref{fig:npp} for
$w=n_0$ in the main plot set, $w=2n_0$ in the top inset, and $w=3n_0$ in the
bottom inset. The resulting expected values are roughly stable at $t=9\,000$ and
equal, respectively, $1.4036$, $1.8192$, and $2.2983$. It is clear from the
figure that, for $w=n_0$, it is the distribution of the sinks' in-degrees that
exerts the greater influence on how the average number of edge-disjoint directed
paths from all sources to one sink is distributed. For $w=3n_0$, it is the
distribution of the non-sink nodes' in-degrees (the mean-$2$ Poisson) that
eventually does it.

\section{The case of an infinite window}

At time $t$, any value of $w$ surpassing $n_0+t-1$ has the effect of an infinite
window; that is, any node in the network may be chosen to receive one of the two
new edges as an incoming edge. When this is the case, none of our conclusions so
far remains valid. Even though the case of infinite $w$ is of little general
interest for modeling real systems (it is inherently dependent on global
properties of the system as a new node comes in), we feel it is worth commenting
on the resulting reach distribution, which differs strikingly from the finite
case [except when $r=1$, since $P_t(1)=n_0/(n_0+t)$ remains of course valid].

\begin{figure}
\centering
\includegraphics[scale=0.45]{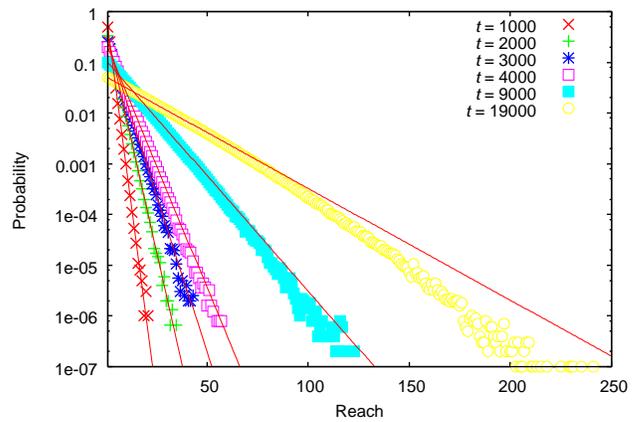}
\caption{(Color online) Reach distribution for $n_0=1\,000$ under an infinite
window. Solid lines give the analytic predictions of (\ref{eq:exp2}). All
simulation data are averages over $500$ independent runs.}
\label{fig:winf}
\end{figure}

Expressing $P_t(r)$ analytically seems infeasible for most values of $r>1$, but
it can be done for $r=2$ and, interestingly, this leads directly to a good
approximation for the general case, provided $t\lessapprox 9n_0$. Notice first
that, for sufficiently large $n_0$,
\begin{eqnarray}
P_t(2)
&\approx&\left(\frac{1}{n_0+t}\right)
\sum_{i=1}^t\left(\frac{n_0}{n_0+i-1}\right)^2\\
&=&P_t(1)n_0\zeta_t(2,n_0),
\end{eqnarray}
where
\begin{equation}
\zeta_t(2,n_0)=\sum_{u=0}^{t-1}\frac{1}{(n_0+u)^2}
\end{equation}
is the truncation, to $t$ terms, of $\zeta(2,n_0)$, Riemann's two-parameter zeta
function \cite{gr00}. Our heuristic generalization for all values of $r$ is
then simply the exponential
\begin{equation}
P_t(r)\approx P_t(1)\left[n_0\zeta_t(2,n_0)\right]^{r-1}.
\label{eq:exp2}
\end{equation}
Simulation results are shown in Figure~\ref{fig:winf}, indicating that, for an
infinite window, reach probabilities fall at least as fast as exponentially.

\section{Discussion and concluding remarks}

We have considered directed networks that grow from a fixed set of ground nodes
by the addition of one node per time step and of two edges directed from that
node to previously existing, randomly chosen nodes inside a fixed-length sliding
window. Networks thus constructed are devoid of directed cycles, and may be
viewed as a prototypical representation of growing collections of partially
ordered items, so long as some underlying time-like notion exists with respect
to which the window mechanism makes sense. Laying down more than two edges per
time step is expected to have no qualitatively significant effect (although it
is unlikely for reaches of small even value to exist in the case of three edges,
for example---a reach of $2$ is in fact impossible---and therefore reach
distributions can be expected to undergo a sort of bifurcation as one moves
from high reaches to lower).

Our study has been centered on the two notions that we deem especially relevant
for the systems acyclic directed networks are purported to relate to. The first
one is the property, here referred to as reachability, of nodes in the network
to be able to reach ground nodes via directed paths. We found, by means of
simulations and also through limited analytic predictions, that the number of
ground nodes reachable from a randomly chosen non-ground node is distributed
first exponentially, then as a power law as time elapses. The other notion on
which we focused can be summarized as that of how to recover a specific ground
node, in the sense of having edge-disjoint directed paths to get to it from some
of the latest nodes to be added to the network. Our finds are that such paths
are expected to occur in very small numbers on average (roughly somewhere near
$2$), and therefore the recoverability of ground nodes may be severely affected
by accidental path disruptions.

We believe this paper's network model, along with its main observables, opens up
new possibilities of investigation about abstract systems that are naturally
representable as acyclic directed networks. Earlier we mentioned examples from
fields related to computer software, artificial intelligence, mathematical
logic, and also history. In addition to their being representable as networks
such as the ones we studied, what these systems also have in common once viewed
from the perspectives of ground-item reachability and recoverability is that
many of them make reference, albeit indirectly, to the growing stack of digital
technologies that currently separates ``ground'' pieces of information from
their representations for end use. Concerns related to this issue are
sometimes voiced in the media, referring, for example, to the digitization of
documents \cite{unesco-url} or to a future in which, as some envisage,
autonomous systems may become inscrutable regarding their internal organization
\cite{itworld-url}. Even though such issues may seem like a far cry from the
study we have pursued in this paper, carrying on with an eye on them may well
prove worthwhile.

\begin{acknowledgments}
The author acknowledges partial support from CNPq, CAPES, and a FAPERJ BBP
grant.
\end{acknowledgments}

\bibliography{dag}
\bibliographystyle{apsrev}

\end{document}